\documentclass[aps,prb,twocolumn,showpacs,preprintnumbers,amsmath,amssymb,superscriptaddress]{revtex4}

\usepackage{graphicx}
\usepackage{dcolumn}
\usepackage{bm}

\begin{document}

\title{Kondo-Dicke resonances in electronic transport through
triple quantum dots}

\author{Piotr Trocha}
\email{piotrtroch@o2.pl}\affiliation{Department of Physics, Adam
Mickiewicz University, 61-614 Pozna\'n, Poland}

\author{J\'ozef Barna\'s}
\email{barnas@amu.edu.pl} \affiliation{Department of Physics, Adam
Mickiewicz University, 61-614 Pozna\'n, Poland}
\affiliation{Institute of Molecular Physics, Polish Academy of
Sciences, 60-179 Pozna\'n, Poland}

\date{\today}

\begin{abstract}
Electronic transport through a triple quantum dot system, with
only a single dot coupled directly to external leads, is
considered theoretically. The model includes Coulomb correlations
in the central dot, while such correlations in the two
side-coupled dots are omitted. The infinite-$U$ mean-field
slave-boson approach is used to obtain basic transport
characteristics in the Kondo regime. When tuning position of the
side-coupled dots' levels, transition from \emph{subradiant} to
\emph{superradiant} like mode (and \emph{vice versa}) has been
found in the spectral function, in analogy to the Dicke effect in
atomic physics. Bias dependence of the differential conductance
and zero frequency shot noise is also analysed.
\end{abstract}

\maketitle

\section{Introduction}
Kondo effect in electronic transport through quantum dots (QDs)
strongly coupled to external leads is a many body phenomenon which
occurs at temperatures $T$ lower than the so-called Kondo
temperature $T_K$, $T<T_K$. The Kondo temperature is
characteristic of a particular system and depends on the
corresponding parameters (energy of the dot level, coupling
strength to external leads, Coulomb parameter). Spin fluctuations
in the dot, generated by coupling of the dot to external leads,
give rise to a narrow peak in the dot's density of states (DOS) at
the Fermi level. This Kondo peak enhances transmission through the
dot, and leads to the unitary limit of the linear conductance at
zero temperature, where the conductance reaches $2e^2/h$. The
Kondo peak in DOS also leads to the so-called zero-bias (Kondo)
anomaly in the differential conductance.

The Kondo effect in electronic transport through single quantum
dots was extensively studied in the last two decades, both
theoretically
\cite{glazman,meir1,meir2,meir3,meir4,kang,dong,swirkowicz,swirkowicz2}
and experimentally \cite{cronewett,gores}. Recently, the
Kondo-assisted transport through double-dot systems in the
T-shaped \cite{wu,sunM,tanaka}, parallel
\cite{tanaka,ding,zhang,sztenkiel,lopez,busser,lim,silva,vernek},
and series \cite{lopez,shon,sunG,aono,aguado,langreth,tanakaS}
geometries also has been addressed. There are, however, only a few
papers on the Kondo phenomenon in electronic transport through
triple-dot structures
\cite{lobos,sun,meisner,kuzmenko,zitko,aldea,tanamoto}. Such
complex dot systems are of current interest from both fundamental
and application points of view. Especially the interference
effects in electronic transport  attract much attention, as the
multi-dot systems offer a unique possibility to study fundamental
phenomena which earlier were observed  in solid state physics
and/or quantum optics. By specific design and fabrication of QD
systems, one can investigate for instance the Aharonov-Bohm
oscillations \cite{wysocki}, Fano resonance \cite{trocha,trocha1},
and others.

Very recently, the Dicke effect, which is well known in atomic
optics \cite{dicke1,dicke2}, has been predicted also for
electronic transport through quantum systems
\cite{shahbazyan,wunsch,guevaraD,orellana1,orellana2,brandes1,brandes2,brandes3}.
The key feature of the Dicke resonance in optics is the presence
of a strong and very narrow spontaneous emission line (in addition
to much broader lines) of a collection of atoms which are
separated by a distance smaller than the wavelength of the emitted
light \cite{dicke1,dicke2}. Generally, the narrow (broad) line is
associated with a state which is weakly (strongly) coupled to the
electromagnetic field, and such long-lived (short-lived) state is
called \emph{subradiant} (\emph{superradiant}) mode. In this paper
we consider another realization of the Dicke-like effect, i.e. the
Kondo-Dicke resonances in electronic transport through a system of
three coupled quantum dots, where two side-coupled dots are
noninteracting (vanishing Coulomb parameter, $U=0$), while the
central dot is coupled to the leads and is in the deep Kondo
regime. In addition, we also observe behavior of the Kondo peak in
DOS, which resembles the transition from \emph{subradiant} to
\emph{superradiant} mode (or \emph{vice versa}) in the usual Dicke
effect. The theoretical approach we use in this paper is based on
the slave-boson technique in the mean-field approximation.

The paper is organized as follows. In section 2 we describe the
model of a three-dot system, and present the corresponding
Hamiltonian in the slave-boson formulation. The mean-field
approach to the problem is described in section 3. Numerical
results on the Kondo problem are shown and discussed in section 4
for symmetric and asymmetric systems. Final conclusions are given
in section 5.

\section{Description of the model}

In this paper we consider a system of three single-level quantum
dots. The whole system is coupled to external leads as shown
schematically in Fig.1. The two side dots, labelled as QD1 and
QD3, are coupled to the central dot QD2 {\it via} a direct hopping
term. The dot QD2, in turn, is additionally attached to external
electron reservoirs (see Fig.1). Moreover, the central dot is
assumed to be in the Kondo regime, while the side dots are beyond
the Kondo  regime (for simplicity they are assumed to be
noninteracting, $U=0$). The system under consideration can be
modelled by the Anderson impurity-like Hamiltonian of the
following form:
\begin{eqnarray}\label{eq1}
\hat{H}=\sum_{\mathbf k\alpha\sigma} \varepsilon_{{\mathbf
k}\alpha \sigma}c^{\dagger}_{{\mathbf k}\alpha \sigma} c_{{\mathbf
k}\alpha \sigma}+
\sum_{i\sigma}\limits\epsilon_{i\sigma}d^\dag_{i\sigma}d_{i\sigma}
   +
   Un_{2\sigma}n_{2\bar{\sigma}}
      \nonumber \\
   + \sum_{j(=1,3)}\sum_{\sigma}\limits(t_{j2\sigma}
  d^\dag_{2\sigma}d_{j\sigma}+\rm h.c.)
  \nonumber \\
   +  \sum_{{\mathbf k}\alpha}\sum_{\sigma}\limits(V_{{\mathbf k}\sigma}^\alpha
   c^\dag_{{\mathbf k}\alpha\sigma}d_{2\sigma}+\rm h.c.).
   \end{eqnarray}
The first term describes the left ($\alpha=L$) and right
($\alpha=R$) electrodes in the non-interacting quasi-particle
approximation, with  $c^{\dagger}_{{\mathbf k}\alpha \sigma}$
($c_{{\mathbf k}\alpha \sigma}$) being the creation (annihilation)
operator of an electron with the wave vector ${\mathbf k}$ and
spin $\sigma$ in the lead $\alpha$, and with
$\varepsilon_{{\mathbf k}\alpha \sigma}$ denoting the
corresponding single-particle energy.

The next three terms of the Hamiltonian describe the system of
three coupled quantum dots. Here, $\epsilon_{i\sigma}$ is the
energy of the discrete level in the $i$-th dot ($i=1,2,3$), while
$t_{j2\sigma}$ denotes the hopping parameter between the $j$-th
dot ($j=1,3$) and the dot QD2. Both $\epsilon_{i\sigma}$ and
$t_{j2\sigma}$ can be spin dependent in a general case.
Furthermore, the term with $U$ describes the intra-dot Coulomb
interactions in the dot QD2, with $U$ denoting the corresponding
Coulomb integral. The inter-dot Coulomb repulsion, similarly as
the intra-dot Coulomb interaction for the dots QD1 and QD3, is
assumed to be small and therefore neglected.

The last term of the system's Hamiltonian describes electron
tunnelling from the leads to the dot QD2 (or {\it vice versa}),
with $V_{{\mathbf k}\sigma}^\alpha$ denoting the relevant matrix
elements ($\alpha =L,R$). Coupling of the dot QD2 to external
leads can be parameterized in terms of
$\Gamma^\alpha_{\sigma}(\varepsilon)=\pi\sum_{\mathbf k}
V_{{\mathbf k}\sigma}^\alpha V^{\alpha\ast}_{{\mathbf
k}\sigma}\delta(\varepsilon-\epsilon_{{\mathbf k}\alpha\sigma})$.
We assume that $\Gamma^\alpha_{\sigma}(\varepsilon )$ is constant
within the electron  band of the leads,
$\Gamma^\alpha_{\sigma}(\varepsilon)=\Gamma^\alpha_{\sigma}={\rm
const}$ for $\varepsilon\in\langle-D,D\rangle$, and
$\Gamma^\alpha_{\sigma}(\varepsilon)=0$ otherwise. Here, $2D$
denotes the electron band width.

\begin{figure}
\begin{center}
\includegraphics[width=0.4\textwidth,angle=0]{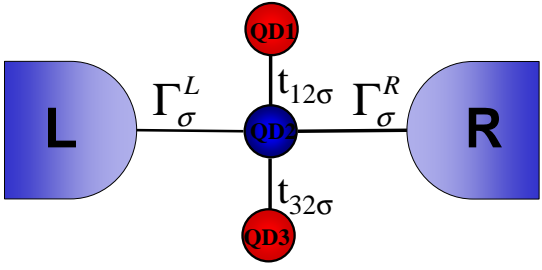}
\caption{Schematic picture of the quantum dot system coupled to
external leads.  $\Gamma_{\sigma}^\alpha$ $(\alpha=L,R)$ describe
coupling of the dot QD2 to the external leads, whereas
$t_{j2\sigma}$ is the hopping parameter between the dot QD2 and
the $j$-th dot, $j=1,3$}
\end{center}
\end{figure}

The following considerations are restricted to the limit of
infinite intra-dot Coulomb parameter for the dot QD2,
$U\rightarrow\infty$, so the double occupancy on the central dot
is forbidden (only one electron can reside in the central dot).
This assumption allows us to employ the infinite-$U$ slave-boson
mean-field (SBMF) approach \cite{coleman} to analyze transport
properties. However, one should bear in mind that the SBMF theory
is valid only for low bias at zero temperature.

In the slave-boson approach, one introduces a set of auxiliary
operators for the central dot. The dot's creation (annihilation)
operator $d^{\dag}_{2\sigma}$ ($d_{2\sigma})$ is then replaced
with $f^{\dag}_{2\sigma}b_2$ ($b_2^{\dag}f_{2\sigma}$). Here,
$b_2$ ($f_{2\sigma}$) is the slave-boson (pseudofermion)
annihilation operator for an empty (singly occupied with a
spin-$\sigma$ electron) state of the dot QD2. To eliminate
non-physical states, the following constraint has to be imposed on
the new quasi-particles,
\begin{equation}\label{}
    \hat{Q}=\sum_{\sigma}\limits
    f^{\dag}_{2\sigma}f_{2\sigma}+b_2^{\dag}b_2=1.
\end{equation}
Equation (2) is a completeness relation of the Hilbert space for
the central dot. Accordingly, the Hamiltonian (1) can be replaced
by an effective Hamiltonian, expressed in terms of the auxiliary
boson $b_2$ and pseudo-fermion $f_{2\sigma}$ operators as
\begin{eqnarray}
\tilde{H}=\sum_{\mathbf k\alpha\sigma} \varepsilon_{{\mathbf
k}\alpha \sigma}c^{\dagger}_{{\mathbf k}\alpha \sigma} c_{{\mathbf
k}\alpha \sigma}+
\sum_{\sigma}\limits\left(\epsilon_{1\sigma}d^\dag_{1\sigma}d_{1\sigma}+
\epsilon_{3\sigma}d^\dag_{3\sigma}d_{3\sigma}\right)
    \nonumber \\
    +
\sum_{\sigma}\limits\epsilon_{2\sigma}f^\dag_{2\sigma}f_{2\sigma}
   +  \frac{1}{\sqrt{N}}\sum_{j(=1,3)}\sum_{\sigma}\limits(t_{j2\sigma}
  f^\dag_{2\sigma}b_2d_{j\sigma}+\rm h.c.)
  \nonumber \\
   +   \frac{1}{\sqrt{N}}\sum_{{\mathbf k}\alpha}\sum_{\sigma}\limits
   (V_{{\mathbf k}\sigma}^\alpha
   c^\dag_{{\mathbf k}\alpha\sigma}b_2^{\dag}f_{2\sigma}+\rm h.c.)
   \nonumber \\
   +\quad\lambda\left(\sum_{\sigma}\limits
    f^{\dag}_{2\sigma}f_{2\sigma}+b_2^{\dag}b_2-1\right).
   \end{eqnarray}
The last term with the Lagrange multiplier $\lambda$ has been
introduced to incorporate the constraint for QD2, given by Eq.(2),
which prevents double occupancy of the central dot. Apart from
this, $N(=2)$ in Eq.(3) stands for the spin-degeneracy of the QD2.
One can check that the number operator $\hat{Q}$ (see Eq.(2))
commutes with this new Hamiltonian, so the total particle number
of $f$-electrons and slave bosons is conserved.

\section{Mean field approach}

The mean field approximation (MFA) adopted to the slave-boson
method relies on replacing the boson field $b_2$ by its
expectation value. Thus, in the lowest order (in the $1/N$
expansion), the slave-boson operator is substituted by a real and
independent of time $c$ number, $b_2(t)/\sqrt{N}\rightarrow\langle
b_2(t)\rangle/\sqrt{N}\equiv{\tilde b}_2$. This approximation
neglects fluctuations around the average value $\langle
b_2(t)\rangle$ of the slave boson operator, but takes into account
spin fluctuations (Kondo regime) -- exactly at $T=0K$ and in the
limit of $N\rightarrow\infty$. On the other hand, one should bear
in mind that MFA cannot be applied in the mixed-valence regime,
where strong charge fluctuations take place. That is why the MFA
approach restricts our considerations to the low bias regime
($eV\ll |\epsilon_2|$). Introducing the following renormalized
parameters $\tilde{t}_{j2\sigma}=\tilde{b}_2t_{j2\sigma}$,
$\tilde{V}_{\mathbf {k}\sigma}^\alpha =V_{\mathbf
{k}\sigma}^\alpha\tilde{b}_2$ and
$\tilde{\epsilon}_{2\sigma}=\epsilon_{2\sigma}+\lambda$, the
effective MFA Hamiltonian can be rewritten as
\begin{eqnarray}
\tilde{H}^{MFA}=\sum_{\mathbf k\alpha\sigma} \varepsilon_{{\mathbf
k}\alpha \sigma}c^{\dagger}_{{\mathbf k}\alpha \sigma} c_{{\mathbf
k}\alpha \sigma}+
\sum_{i=1,3}\sum_{\sigma}\limits\epsilon_{i\sigma}d^\dag_{i\sigma}d_{i\sigma}
    \nonumber \\
    +
\sum_{\sigma}\limits\tilde{\epsilon}_{2\sigma}f^\dag_{2\sigma}f_{2\sigma}
   +  \sum_{j(=1,3)}\sum_{\sigma}\limits(\tilde{t}_{j2\sigma}
  f^\dag_{2\sigma}d_{j\sigma}+\rm h.c.)
  \nonumber \\
   +   \sum_{{\mathbf k}\alpha}\sum_{\sigma}\limits
   (\tilde{V}_{\mathbf{k}\sigma}^\alpha
   c^\dag_{{\mathbf k}\alpha\sigma}f_{2\sigma}+{\rm h.c.})
  +  \lambda\left(
    N\tilde{b}^2-1\right).
   \end{eqnarray}

One needs now to find the unknown parameters $\tilde{b}_2$ and
$\lambda$. These can be determined from the constraint imposed on
the slave bosons (Eq.(2)), and from the equation of motion for the
slave boson operators. As a result, one finds two equations,
\begin{equation}\label{}
    \frac{1}{N}\sum_{\sigma}\limits
    \langle
    f^{\dag}_{2\sigma}f_{2\sigma}\rangle+\tilde{b}_2=\frac{1}{N},
\end{equation}

\begin{equation}\label{}
    \sum_{j=1,3}\sum_{\sigma}\limits\tilde{t}_{j2\sigma}^\ast\langle
    d^{\dag}_{j\sigma}f_{2\sigma}\rangle+
\sum_{{\bf
k}\alpha\sigma}\limits\tilde{V}_{\mathbf{k}\sigma}^\alpha\langle
    c^{\dag}_{{\bf k}\alpha\sigma}f_{2\sigma}\rangle+
    N\lambda\tilde{b}_2=0.
\end{equation}

There is also another way to obtain the above equations. The free
parameters $\tilde{b}_2$ and $\lambda$ can be determined by
minimizing the ground state energy of the effective MFA
Hamiltonian with respect to these parameters. Application of the
Hellmann-Feynman theorem to the Hamiltonian (4), together with the
conditions for minimal energy
$({\partial\langle\tilde{H}^{MFA}\rangle}/{\partial\tilde{b}})=0=
({\partial\langle\tilde{H}^{MFA}\rangle}/{\partial\lambda})$,
gives a set of self-consistent equations. These equations in the
Fourier space read
\begin{equation}\label{}
    \tilde{b}_2-i\sum_{\sigma}\limits
    \int\frac{d\varepsilon}{2\pi N}\langle\langle
    f_{2\sigma}|f_{2\sigma}^{\dag}\rangle\rangle^<_\varepsilon =\frac{1}{N},
\end{equation}

\begin{eqnarray}\label{}
-i\sum_{j=1,3}\sum_{\sigma}\limits\tilde{t}_{j2\sigma}^\ast\int\frac{d\varepsilon}{2\pi}
    \langle\langle
    f_{2\sigma}|d^{\dag}_{j\sigma}\rangle\rangle^<_\varepsilon\qquad\qquad\qquad\quad
\nonumber\\
    -i
\sum_{{\bf
k}\alpha\sigma}\limits\tilde{V}_{\mathbf{k}\sigma}^\alpha\int\frac{d\varepsilon}{2\pi}\langle\langle
    f_{2\sigma}|c^{\dag}_{{\bf k}\alpha\sigma}\rangle\rangle^<_\varepsilon +
    N\lambda\tilde{b}_2=0,
\end{eqnarray}
where $\langle\langle
    f_{2\sigma}|d^{\dag}_{j\sigma}\rangle\rangle^<_\varepsilon$ and $\langle\langle
    f_{2\sigma}|c^{\dag}_{{\bf k}\alpha\sigma}\rangle\rangle^<_\varepsilon$
    are the Fourier transforms of the
    lesser Green functions; the former being defined as $G^<_{2j}(t,t^\prime)\equiv \langle\langle
    f_{2\sigma}(t)|d^{\dag}_{j\sigma}(t')\rangle\rangle^<=i
    \langle
    d^{\dag}_{j\sigma}(t')f_{2\sigma}(t)\rangle$ and a similar definition also holds for
    $\langle\langle
    f_{2\sigma}(t)|c^{\dag}_{{\bf k}\alpha\sigma}(t^\prime)\rangle\rangle^<$.

    To rewrite Eq.(8) in a different form, we write first the equation of
    motion for  $f_{2\sigma}(t)$. Then, we
    multiply this equation by $f_{2\sigma}^\dag(t')$ and take
    the thermal average. Upon expressing the obtained equation in terms of
    the lesser Green functions, and taking its hermitian conjugation, one
    finds
\begin{eqnarray}\label{}
(\varepsilon-\tilde{\epsilon}_{2\sigma})\langle\langle
    f_{2\sigma}|f^{\dag}_{2\sigma}\rangle\rangle^<_\varepsilon =
    \sum_{j=1,3}\sum_{\sigma}\limits\tilde{t}_{j2\sigma}^\ast
    \langle\langle
    f_{2\sigma}|d^{\dag}_{j\sigma}\rangle\rangle^<_\varepsilon
    \nonumber
    \\
    +\quad\sum_{{\bf
k}\alpha\sigma}\limits\tilde{V}_{\mathbf{k}\sigma}^\alpha\langle\langle
    f_{2\sigma}|c^{\dag}_{{\bf k}\alpha\sigma}\rangle\rangle^<_\varepsilon .
\end{eqnarray}
When substituting Eq.(9) into Eq.(8), one finally arrives at the
equation
\begin{eqnarray}\label{}
-i\sum_{\sigma}\limits\int\frac{d\varepsilon}{2\pi}(\varepsilon-\tilde{\epsilon}_{2\sigma})
\langle\langle
f_{2\sigma}|f^{\dag}_{2\sigma}\rangle\rangle^<_\varepsilon +
N\lambda\tilde{b}_2=0.
\end{eqnarray}

Formulas (7) and (10) are our final equations for the unknown
parameters. To solve them one still needs to determine the lesser
Green function $\langle\langle
f_{2\sigma}|f^{\dag}_{2\sigma}\rangle\rangle^<_\varepsilon$. This
can be derived from the corresponding equation of motion.
Similarly, the equation of motion can also be used to derive the
retarded Green function (which is also required in our
calculations). As a result one finds
\begin{equation}\label{}
    G^r_{22\sigma}(\varepsilon)=\frac{1}
    {\varepsilon-\tilde{\epsilon}_{2\sigma}+i\tilde{\Gamma}_{\sigma}-
    \frac{\tilde{t}_{32\sigma}^2}{\varepsilon-\epsilon_{3\sigma}}-
    \frac{\tilde{t}_{12\sigma}^2}{\varepsilon-\epsilon_{1\sigma}}},
\end{equation}
\begin{equation}\label{}
    G^<_{22\sigma}(\varepsilon)=\frac{2i[f_L(\varepsilon)\tilde{\Gamma}_{\sigma}^L+
    f_R(\varepsilon)\tilde{\Gamma}_{\sigma}^R]}
    {\left|\varepsilon-\tilde{\epsilon}_{2\sigma}+i\tilde{\Gamma}_{\sigma}-
    \frac{\tilde{t}_{32\sigma}^2}{\varepsilon-\epsilon_{3\sigma}}-
    \frac{\tilde{t}_{12\sigma}^2}{\varepsilon-\epsilon_{1\sigma}}\right|^2},
\end{equation}
where $f_\alpha(\varepsilon)$ is the Fermi-Dirac distribution
function in lead $\alpha$,
$\tilde{\Gamma}_{\sigma}^{\alpha}=\tilde{b}_{2}^2\Gamma_{\sigma}^{\alpha}$
and $\tilde{\Gamma}_{\sigma}=\tilde{\Gamma}_{\sigma}^{L}+
\tilde{\Gamma}_{\sigma}^{R}$ are the renormalized parameters
describing coupling of the dot QD2 to external leads.

Electric current $J$ flowing through the system can be determined
from the formula
\begin{equation}\label{}
    J=\frac{e}{h}\sum_{\sigma}\int d\varepsilon
    [f_L(\varepsilon)-f_R(\varepsilon)]T_{\sigma}(\varepsilon),
\end{equation}
where
$T_{\sigma}(\varepsilon)=4G_{22\sigma}^a\tilde{\Gamma}_{\sigma}^{R}
G_{22\sigma}^r\tilde{\Gamma}_{\sigma}^{L}$ is the transmission
probability. It is worth noting that the above current formula is
similar to that used for noninteracting QD systems. In our case,
however, the transmission probability depends on the renormalized
parameters $\tilde{\Gamma}_{\sigma}^{\alpha}$,
$\tilde{\epsilon}_{2\sigma}$ and $\tilde{t}_{j2\sigma}$, which
depend on the gate and transport voltages. At $T=0{\mathrm K}$,
the current and linear conductance are given by the formula
\begin{equation}\label{}
    J_{0}=\frac{e}{h}\sum_{\sigma}\int_{-eV/2}^{eV/2}d\varepsilon\;
    T_{\sigma}(\varepsilon),
\end{equation}
and
\begin{equation}\label{}
    G_{V\to 0}=\lim_{V\rightarrow 0}\frac{dJ}{dV}=\frac{e^2}{h}\sum_\sigma
    T_\sigma(\varepsilon=0).
\end{equation}

\section{Numerical results}

In numerical calculations we assume that the dot levels and the
hoping parameters are independent of electron spin,
$\epsilon_{i\sigma}=\epsilon_i$ (for $i =1,2,3$),
$t_{j2\sigma}=t_{j2}$ for $j=1,3$, and $\Gamma_\sigma^\alpha
=\Gamma^\alpha$ (for $\alpha =L,R$). The energy levels
$\epsilon_i$ are measured from the Fermi level of the leads in
equilibrium ($\mu_L=\mu_R=0$). In the following we set the bare
level of the dot QD2 at $\epsilon_{2}=-3.5\Gamma$, and the
bandwidth is assumed to be $D=60\Gamma$. In this paper all the
energy quantities will be expressed in the units of $\Gamma$
($\Gamma=\Gamma^L+\Gamma^R$). The energy levels $\epsilon_1$ and
$\epsilon_3$ can be tuned by applying gate voltages to QD1 and
QD3. Taking into account the above parameters, the Kondo
temperature $T_K$ of the central dot  for $t_{12}=t_{32}=0$ can be
estimated to be $T_K=10^{-3}\Gamma$ ($k_B=1$).

\begin{figure}
\begin{center}
\includegraphics[width=0.46\textwidth]{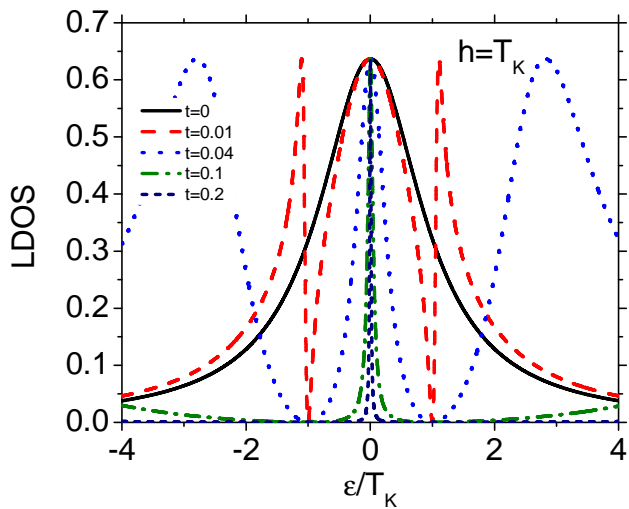}
\caption{Local density of states for the dot QD2, calculated for
indicated values (in the units of $\Gamma$) of the interdot
coupling strength $t_{12}=t_{32}=t$, and for
$\epsilon_1=-\epsilon_3\equiv h=T_K=0.001\Gamma$.}
\end{center}
\end{figure}

\begin{figure}
\begin{center}
\includegraphics[width=0.46\textwidth]{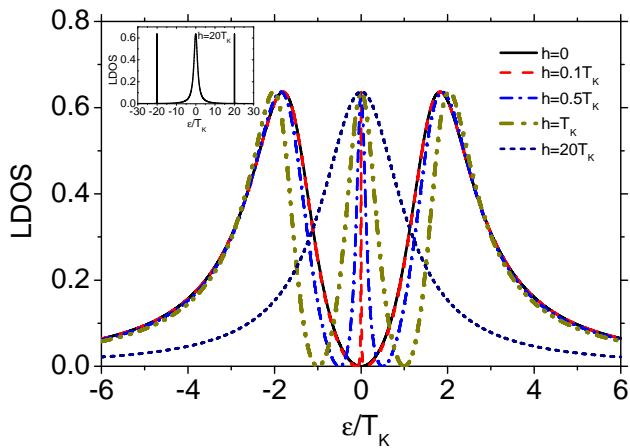}
\caption{Local density of states for the dot QD2 calculated for
indicated values of $h$ ($h=\epsilon_1=-\epsilon_3$),
 and for the interdot coupling
$t_{12}=t_{32}=t=0.03\Gamma$.}
\end{center}
\end{figure}

\subsection{Density of states}

We start from the local density of states (LDOS) at equilibrium,
$D_2$, for the central dot QD2, which can be calculated from the
formula
\begin{equation}\label{}
   D_2=-\frac{\tilde{b}_2^2}{\pi}\sum_{\sigma}\Im\left[G_{22\sigma}^r(\varepsilon)\right],
\end{equation}
where $\Im[A]$ denotes the imaginary part of $A$.

To understand basic features of the LDOS $D_2$ in the Kondo regime
and the influence of side-coupled dots, we consider first the
situation when the dot levels are fixed while the coupling between
the dots can be changed. The corresponding LDOS  is shown in Fig.2
for several values of $t_{12}=t_{32}\equiv t$. Assume first that
the central dot is detached from the two dots QD1 and QD3, $t=0$
(solid line in Fig.2). The peak at the Fermi level in $D_2$
reveals then the usual Kondo phenomenon in a single dot, which
leads to the Kondo anomaly in transport through the dot (to be
discussed later). If now one additional and noninteracting dot is
attached to the Kondo dot {\it via} a hopping term, the shape of
the Kondo peak is changed due to the interference effects (not
shown in Fig.2). As it has been shown by Wu \emph{et al}
\cite{wu}, the DOS can be then decomposed into a Breit-Wigner and
a Fano line shape. Assume now that another non-interacting dot has
been connected to the Kondo dot, as in the case shown in Fig.1.
Let the bare levels of the side-coupled dots are fixed at $T_K$
and $-T_K$, respectively. When the interdot coupling strength $t$
is nonzero, two additional asymmetric satellite peaks emerge in
LDOS $D_2$. As shown in Fig.2, these peaks move away from the
central peak as $t$ increases. At the same time, the widths of the
satellite peaks increase, while the central peak becomes narrower.

In Fig.3 the LDOS for the dot QD2 is shown for different energy
levels of the side-coupled dots, while coupling of these dots to
the central dot is fixed, $t_{12}=t_{32}=t=0.03\Gamma$. It has
been assumed there that the energy levels of the dots QD1 and QD3
are located symmetrically with respect to the Fermi level at
equilibrium, $\epsilon_1=h$ and $\epsilon_3=-h$. For a nonzero
$h$, the LDOS reveals three well defined peaks. The central peak
(the main Kondo peak) is located at the Fermi level of the leads,
while the other two satellite peaks are roughly at
$\pm\sqrt{h^2+2\tilde{t}^2}$. When $h<<t$, position of these
satellite peaks is only weakly dependent on $h$. Between the
maxima, the LDOS goes to zero at $\epsilon=\pm h$. The width of
central peak becomes narrower and narrower with decreasing $h$,
and the peak has almost $\delta$-like shape for very small values
of  $h$, $h\ll t$. This means that the Kondo state associated with
this maximum becomes then more and more localized at the Fermi
level. In analogy to the Dicke effect, which is well known in
optics, the narrow central peak in LDOS may be considered as a
long-lived (\emph{subradiant}) state, whereas the other two peaks
as corresponding to short-lived (\emph{superradiant}) states. It
is interesting to note that one can observe transition from
\emph{subradiant} to \emph{superradiant} mode (and \emph{vice
versa}) by tuning separation of the side-coupled dots' levels from
the Fermi level, e.g. by gate voltages. Thus, a particular state,
for example the one at the Fermi level, is \emph{subradiant}-like
for small values of $h$ ($h\ll t$), and \emph{superradiant}-like
for large values of $h$ ($h\gg t$). When the energy levels of the
side-coupled dots coincide with the Fermi level, $h=0$, the Kondo
peak (central peak) in LDOS is suppressed and disappears.

\begin{figure}
\begin{center}
\includegraphics[width=0.46\textwidth]{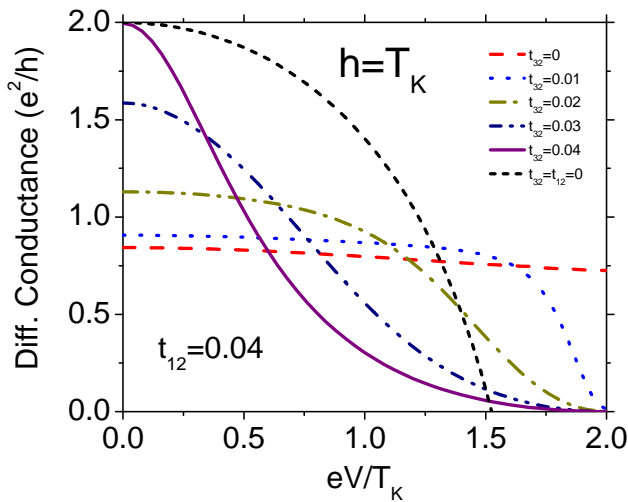}
\caption{Differential conductance as a function of the applied
bias voltage for indicated values of the interdot couplings
(measured in the units of $\Gamma$).}
\end{center}
\end{figure}

For large values of the side dots' level separation ($h\gg t$),
the satellite peaks are narrow, while the central peak is then
relatively broad. This situation is qualitatively similar to that
observed in Ref.[\onlinecite{trochaJP}]. However, the width of the
central peak is now a complex non-monotonic function of the level
separation $h$ (due to self-consistent parameters $\tilde{b}_2$,
$\lambda$). The width of the peak increases for small values of
$h$ achieving a maximum and then decreases very slowly (even for
large values of $h$). Moreover, in the nonlinear response regime,
when a nonzero bias voltage is applied to the system, the width of
the central peak is further suppressed, especially for large $h$.
Position of the maximum moves to smaller values of $h$ as bias
voltage increases. Another new feature is  a broad range of very
small LDOS between the central maximum and the satellite peaks for
larger values of $h$.

To summarize this section we conclude that the width of the Kondo
peak in LDOS can be changed by tuning both the level position of
side dots  and the inter dot coupling strength. Such systems give
us an opportunity to study cross-over from the \emph{subradiant}
to \emph{superradiant} mode, and \emph{vice versa}.

\begin{figure}
\begin{center}
\includegraphics[width=0.46\textwidth]{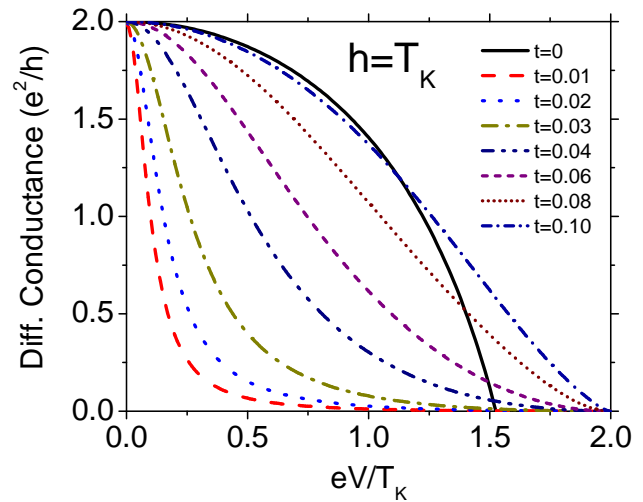}
\caption{Differential conductance as a function of the bias
voltage for indicated values of the interdot couplings
$t=t_{12}=t_{32}$ (measured in the units of $\Gamma$).}
\end{center}
\end{figure}

\begin{figure}
\begin{center}
\includegraphics[width=0.46\textwidth]{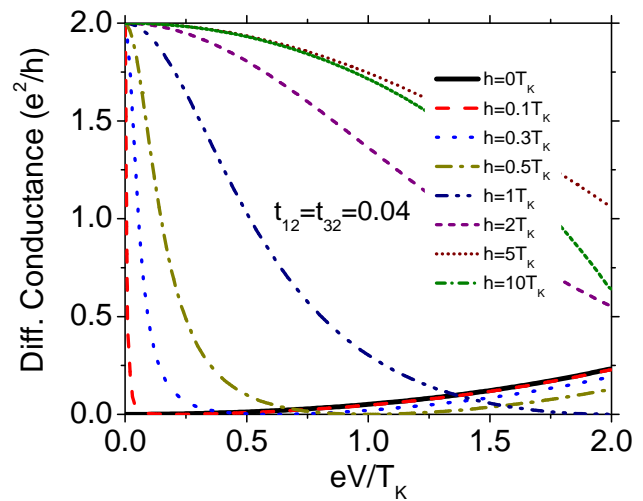}
\caption{Differential conductance as a function of the bias
voltage for indicated value od the coupling strengths (measured in
the units of $\Gamma$), and indicated energy levels of the
side-coupled dots, $\epsilon_1=-\epsilon_3=h$.}
\end{center}
\end{figure}

\subsection{Differential conductance}

Now we consider nonlinear differential conductance $dJ/dV$. In
Fig.4 we show differential conductance as a function of the
applied voltage for the case when the dot QD1 is coupled to the
Kondo dot with a constant strength, $t_{12}=0.04\Gamma$, while the
coupling strength of the dot QD3 is gradually switched on. For
comparison we have also shown there the curve for the Kondo dot
totally decoupled from both side dots, $t_{12}=t_{32}=0$. Owing to
the Kondo resonance, the unitary limit of the conductance is then
reached for zero bias limit. This is well know behavior of
transport through a single Kondo dot. Let us now turn on the
coupling of the Kondo dot to one of the side dots, say to the dot
QD1. The zero bias anomaly becomes then suppressed as a
consequence of the interplay between the Fano interference and the
Kondo effect \cite{wu}. For a sufficiently large $t_{12}$, the
zero bias anomaly may even disappear. However, the situation
becomes more complex when the second side dot (QD3) is attached to
the Kondo dot and starts to play a role in transport. With
increasing coupling strength $t_{32}$ of the dot QD3 to the dot
QD2, one observes revival of the zero bias anomaly peak, which for
$t_{32}=t_{12}$ reaches unitary limit for zero bias. Interplay of
the Dicke and Kondo resonances leads, however, to a faster
decrease of the conductance at small voltages, and slower at
higher voltages. As a consequence, a tail in the conductance
survives at higher voltages, where the conductance of the Kondo
dot detached from the side dots already disappears. This behavior
is a consequence of the narrowing of the central Kondo peak in
LDOS and the occurrence of two satellite peaks (see discussion in
the previous subsection). This behavior is clearly seen in Fig. 5,
where the bias dependence of the differential conductance is shown
for several values of the inter-dot coupling parameters,
$t_{32}=t_{12}=t$. The zero bias conductance  in such a
symmetrical case does not depend on the value of $t$, which
reflects the LDOS behavior (see Fig.2).

Similar behavior also occurs when the coupling strength
$t_{12}=t_{32}\equiv t$ are fixed while the level separation of
the side-coupled dots is changed (see Fig.6). For large values of
$h$ the conductance decreases with increasing bias monotonically.
However, for small values of $h$ the differential conductance is a
nonmonotonic function of the applied bias voltage. It falls down
rapidly in the region of small bias voltages achieving a minimum
value at some voltage, and then rises again. These features are
quite reasonable because the satellite peaks move away from the
central peak as $h$ increases. Thus, for large values of $h$ the
satellite peaks do not  contribute to electronic transport. For a
small $h$, however, the satellite peaks enter the transport window
and contribute to current, leading to an increase in the
conductance above a certain bias voltage. It is worth to note,
that the zero bias anomaly for $h=0$ is totally suppressed and the
differential conductance slowly rises as the bias voltage is
applied. This suppression is a consequence of the suppression of
the Kondo peak in DOS due to the interplay of the Dicke and Kondo
resonances.

\begin{figure}
\begin{center}
\includegraphics[width=0.48\textwidth]{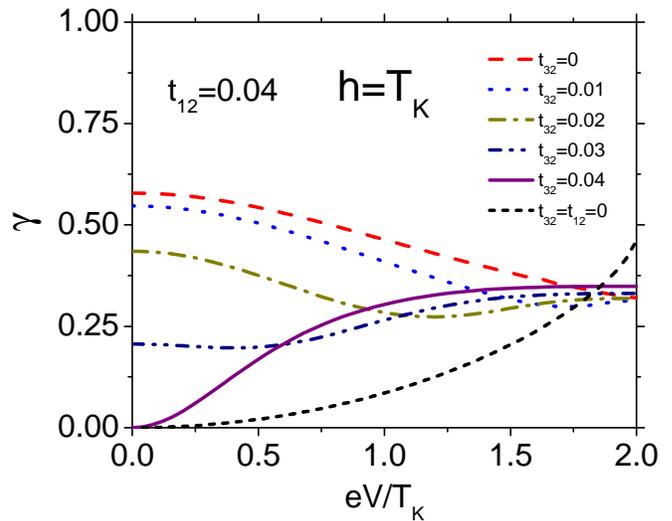}
\caption{Fano factor as a function of the applied bias voltage for
indicated values of the interdot couplings (measured in the units
of $\Gamma$).}
\end{center}
\end{figure}

\subsection{Shot noise}

The only nonzero contribution to the noise in electronic transport
at zero temperature, $T=0K$, is the shot noise, while the thermal
noise is then totally suppressed. The quadratic form of the MFA
Hamiltonian allows us to express the zero frequency shot noise in
the form,
\begin{equation}\label{}
    S=\frac{2e^2}{h}\int_{-eV/2}^{eV/2}d\epsilon
    T_{\sigma}(\epsilon)[1-T_{\sigma}(\epsilon)].
\end{equation}
Deviation of the shot noise $S$ from the Poisson value, $S_P=2eJ$,
is usually characterized by the Fano factor $\gamma=S/2eJ$, which
for the Poissonian noise is exactly equal to 1.

In Fig. 7, the bias dependence of the Fano factor is displayed for
different interdot coupling parameters $t_{32}$, whereas the
parameter $t_{12}$ is fixed at $0.04\Gamma$. For the T-shape
geometry (one side dot is decoupled from the central dot QD2,
$t_{32}=0$), the Fano interference plays a significant role in
transport processes, and -- as discussed above -- leads to
suppression of the zero bias anomaly below the unitary limit. This
is also manifested in the shot noise characteristics. The zero
bias limit of the Fano factor is larger than zero, and this
behavior is a result of the weakening of the Kondo peak at the
Fermi level due to the Fano interference. With increasing
$t_{12}$, but still having $t_{32}=0$, the Fano interference
outweighs the Kondo effect and the Fano factor $\gamma$ tends to
unitary limit (shot noise tends to the Poissonian value). When
coupling of the dot QD3 to the dot QD2 is turned on, the zero bias
Fano factor drops and finally for a symmetric system,
$t_{12}=t_{32}$, reaches zero as for the single quantum dot. As a
result, the system is ideally transparent for electrons at the
Fermi energy $\mu=0$ (at the equilibrium).
\begin{figure}
\begin{center}
\includegraphics[width=0.48\textwidth]{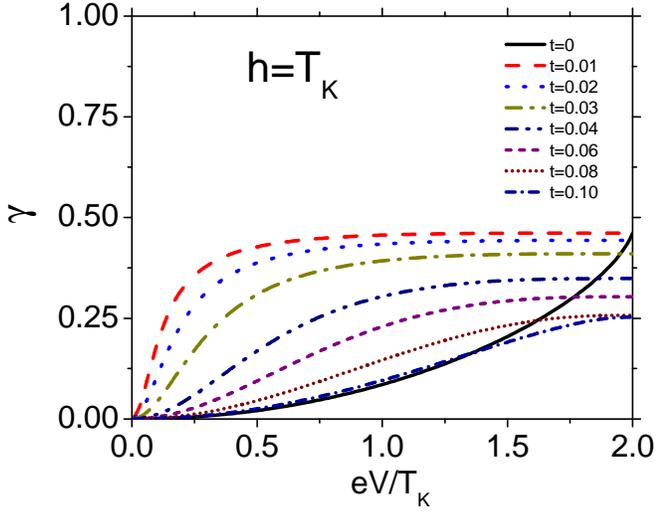}
\caption{Fano factor as a function of the bias voltage for various
values of the interdot couplings $t=t_{12}=t_{32}$ (measured in
the units of $\Gamma$).}
\end{center}
\end{figure}

Figure 8 shows the Fano factor as a function of the applied
voltage for various inter-dot couplings, $t_{32}=t_{12}=t$.
Applied bias voltage partially suppresses the Kondo effect, which
results in the growth of the Fano factor. This growth is faster
for smaller values of $t$.

\begin{figure}
\begin{center}
\includegraphics[width=0.48\textwidth]{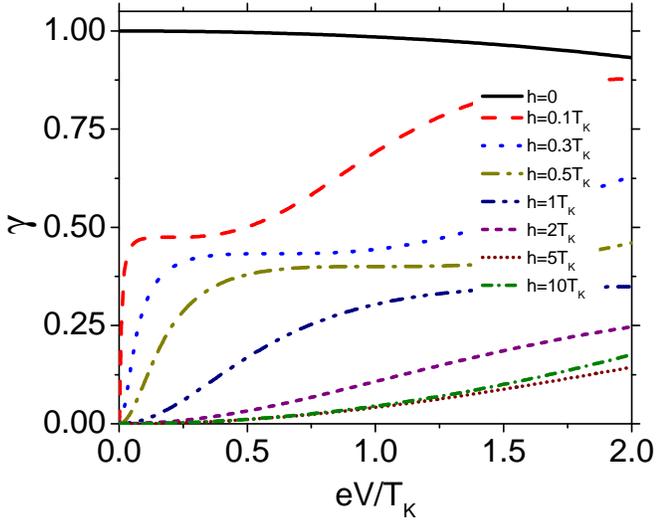}
\caption{Fano factor as a function of the bias voltage for
indicated side-coupled dots' levels; $\epsilon_1=h$,
$\epsilon_3=-h$, and for $t_{12}=t_{31}=0.04\Gamma$.}
\end{center}
\end{figure}

\begin{figure}
\begin{center}
\includegraphics[width=0.48\textwidth]{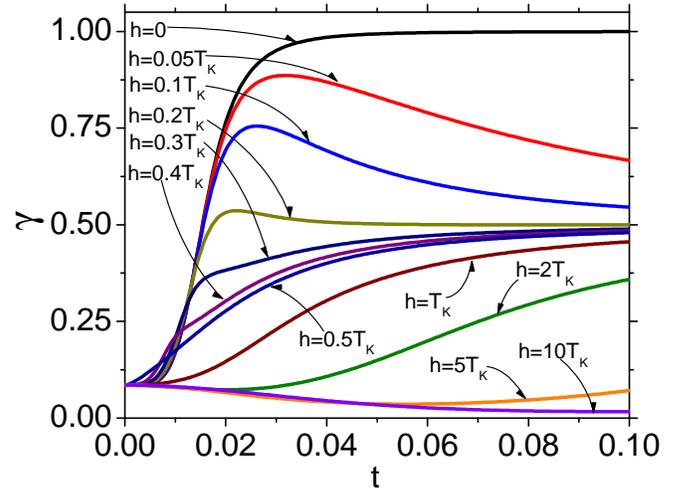}
\caption{Fano factor as a function of the interdot coupling
strength $t$ for different values of $h$; $\epsilon_1=h$,
$\epsilon_3=-h$. The bias is fixed at $eV=T_K$.}
\end{center}
\end{figure}

When the levels of the side coupled dots are located at the Fermi
level ($h=0$), the zero bias Fano factor achieves the Poissonian
value and then slowly falls down with increasing bias voltage (see
Fig.9). In contrary, the zero bias limit of the Fano factor for a
nonzero level separation of the side-coupled dots is zero and
tends to the Poissonian limit with increasing bias voltage. One
can then note a sudden increase of the Fano factor for a small
value of the level separation, $h<T_K$, while for $h>T_K$ the Fano
factor remains small over a wider range of the bias voltage before
it grows to the Poisson limit.

In Fig.10 the Fano factor $\gamma$ is plotted as a function of $t$
for a fixed bias voltage $eV=T_K$, and  for different positions of
the side-coupled dots' levels ($\epsilon_1=h$, $\epsilon_3=-h$).
When the bare levels of side-coupled dots are in the transport
window, the Fano factor is then significantly affected by the
interdot coupling. For a very small value of $h$, the Fano factor
grows rapidly from a small value at zero interdot coupling to its
maximum values, and then falls down to half of the Poissonian
value at sufficiently strong couplings. It is worth to note that
the Fano factor does not reach the Poissonian limit for any $h$,
except the case of $h=0$. In turn, for $h>>T_K$ the Fano factor
$\gamma$ is less sensitive to the interdot coupling strength. The
central peak in the LDOS is then very broad and gives a large
contribution to the differential conductance. On can also note
that for a large value of $h$, $\gamma$ remains small in a wider
range of the interdot coupling before it rises to half of the
Poisson value. Contrary to the case of small $h$, one can now see
a minimum in the the Fano factor (for a finite bias). Thus, the
Kondo-assisted transport is optimized for $t=t_{min}$.

\section{Summary and conclusions}

In this paper we have considered the effects due to interplay of
the  Kondo and Dicke resonances in electronic transport through a
three-dot system. One of the dot (that attached to the leads) was
in the Kondo regime, while the two side dots were out of the Kondo
regime. Using the method based on the slave boson mean field
approximation we have calculated local density of states,
differential conductance, and shot noise. The results clearly show
that the side dots strongly modify the transport characteristics.

The Kondo peak in the local density of states of the central dot
becomes narrower due to the interplay of the Kondo and Dicke
effects. The Kondo peak is even totally suppressed when the bare
energy levels of the side dots are equal and located at the Fermi
level. This leads to suppression of the zero bias Kondo peak in
the differential conductance. The interplay of the Kondo an Dicke
resonances has also a significant impact on the shot noise
characteristics.

\begin{acknowledgements}
This work, as part of the European Science Foundation EUROCORES
Programme SPINTRA, was supported by funds from the Ministry of
Science and Higher Education as a research project in years
2006-2009 and the EC Sixth Framework Programme, under Contract N.
ERAS-CT-2003-980409.

\end{acknowledgements}

\end{document}